\documentclass[twocolumn,superscriptaddress,showkeys]{revtex4-1}

\usepackage{xcolor}
\usepackage{amsfonts}
\usepackage{graphicx}
\usepackage{bm}
\usepackage{dcolumn}
\usepackage[colorlinks,linkcolor=blue,citecolor=blue]{hyperref}
\usepackage{multirow}
\usepackage{amsmath}
\usepackage{hhline}
\usepackage{mathptmx}
\usepackage{eucal}
\usepackage{color}
\usepackage{epstopdf}
\usepackage{natbib}
\usepackage{textcomp}

\usepackage[T1]{fontenc}
\usepackage[latin9]{inputenc}
\usepackage{amsmath,amssymb,amstext}
\usepackage{esint}
\usepackage{bm,bbold}
\usepackage{ulem}

\begin{document}


\title[]{A superconducting absolute spin valve}

\author{G. De Simoni}
\email{giorgio.desimoni@sns.it}
\affiliation{NEST Istituto Nanoscienze-CNR  and Scuola Normale Superiore, I-56127 Pisa, Italy}
\author{E. Strambini}
\affiliation{NEST Istituto Nanoscienze-CNR  and Scuola Normale Superiore, I-56127 Pisa, Italy}
\author{J. S. Moodera}
\affiliation{Department of Physics, Francis Bitter Magnet Lab and Plasma Science and Fusion Center, Massachusetts Institute of Technology, Cambridge, Massachusetts 02139, USA}
\author{F. S. Bergeret}
\email{sebastian_bergeret@ehu.eus}
\affiliation{Centro de Fisica de Materiales (CFM-MPC), Centro Mixto CSIC-UPV/EHU,Manuel de Lardizabal 5, E-20018 San Sebastian, Spain}
\affiliation{Donostia International Physics Center (DIPC), Manuel de Lardizabal 4, E-20018 San Sebastian, Spain}
\author{F. Giazotto}
\email{francesco.giazotto@sns.it}
\affiliation{NEST Istituto Nanoscienze-CNR  and Scuola Normale Superiore, I-56127 Pisa, Italy}


\maketitle

\textbf{
A superconductor with a spin-split excitation spectrum behaves as an ideal ferromagnetic spin-injector in a tunneling junction\cite{DeGennes1966,meservey1970magnetic,tedrow1971spin,Giazotto2008}. It was theoretical predicted that the combination of two  such spin-split superconductors with independently tunable magnetizations, may be used as an ideal \textit{absolute} spin-valve\cite{Huertas-Hernando2002}. 
  Here we report on the first  switchable superconducting spin-valve based on two EuS/Al bilayers coupled through an aluminum oxide tunnel barrier. The spin-valve shows a relative resistance change between the parallel and antiparallel configuration of the EuS layers up to 900\% that demonstrates a  highly spin-polarized currents through the junction.  Our device  may be pivotal for 
 realization of thermoelectric radiation detectors \cite{ozaeta2014predicted,Giazotto2015} ,   logical element for a memory cell  in cryogenics superconductor-based computers and  superconducting spintronics in general\cite{Linder2015}}.

Data storage is one of the application fields in which spintronics has emerged as a breakthrough technology\cite{Baibich1988,Moodera1995,Chappert2007}:
contemporary computer hard drives and nonvolatile  magnetic  random  access  memories, exploit spin-polarized electron tunneling through magnetic junction device whose functionality  relies upon an efficient control of their electrical tunnel magneto-resistance (TMR)\cite{Moodera1995,Moodera2007,Moodera2010}.
In general terms the underlying physics can be illustrated by the \textit{two current} model\cite{Mott1964},in which  the electronic transport is described as the  parallel of two current channels with opposite spins. In a ferromagnet these two conduction channels have different conductivity and hence spin-polarized currents can be generated by electric means.

Much of the standard technology for  magnetic memories is founded on such spin-polarized currents, which are  usually  created in stacks of two ferromagnetic layers separated either by a non-magnetic metal or a tunneling barrier\cite{Chappert2007}, a structure called \textit{spin-valve}.  
The switching between a parallel (P) and an antiparallel (AP) configuration of the magnetizations of the ferromagnetic layers results in two states with different conductances that encode the logical-0 and logical-1.  A measurement of the spin valve efficiency is given by the relative resistance change between the AP and P configurations. In the case that the junction is made of two ideal ferromagnets with full spin-polarization (half-metallic limit) the AP configuration  corresponds to a zero current state, whereas a finite current flows in the P configuration.
\begin{figure}[th!]
\includegraphics[width=0.9 \columnwidth]{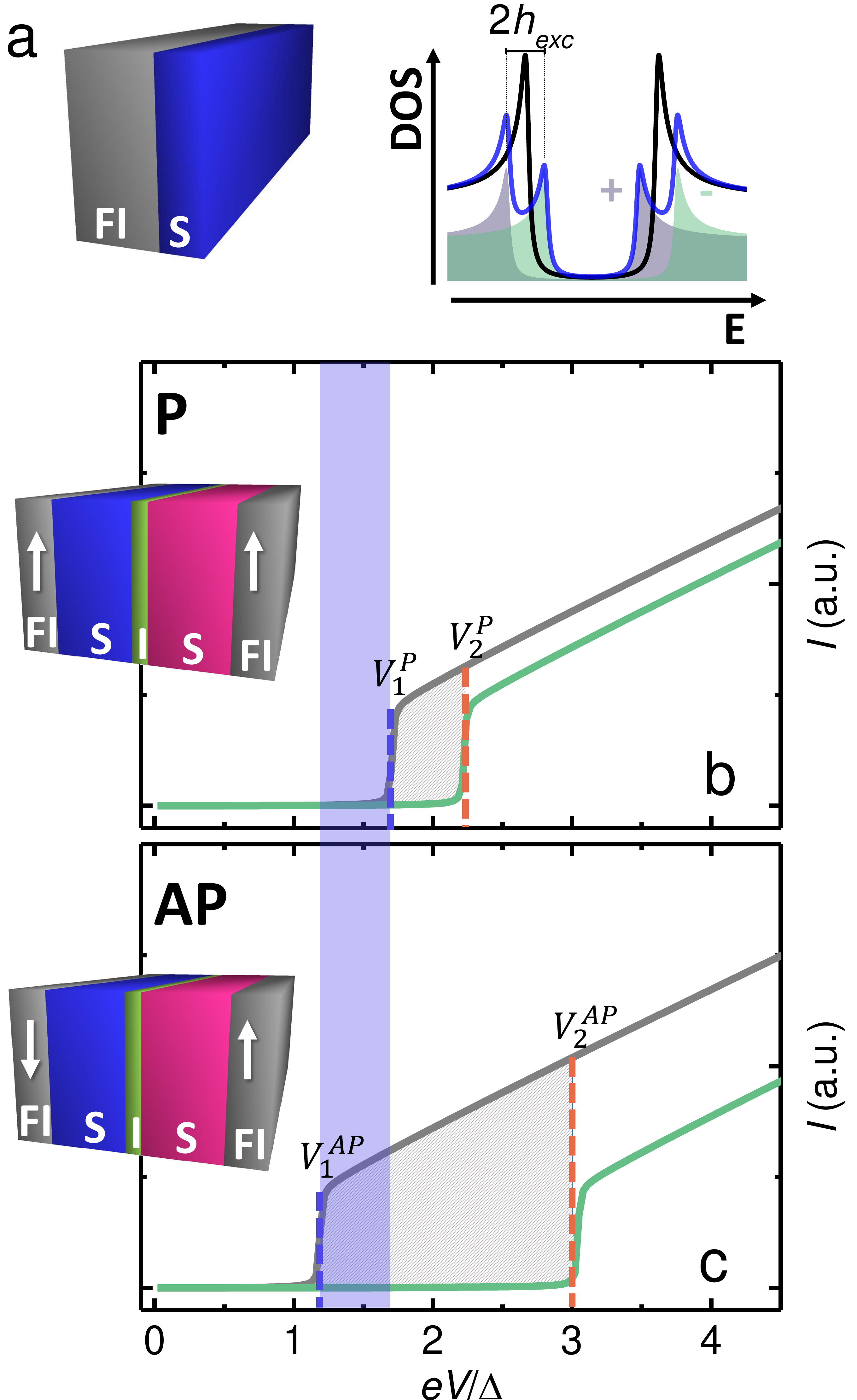}
\caption{\label{fig1} \textbf{Working principle of the absolute spin valve}. \textbf{a,} A ferromagnetic insulator (FI, gray block) in contact with a superconductor (S, blue block) induces through exchange interaction an energy splitting in the density of states (DOS, black curve). This results in double-peaked DOS (blue curve) in which the spin degeneracy is lifted. DOS for $+$ (gray area) and $-$ (green area) spin populations are also  shown.  \textbf{b,} Tunneling currents $I_+$ (grey) and $I_-$ (green) as a function of the voltage bias calculated for the two spin species independently by means of formulas $(1)$ and $(2)$ for a FI-S-I-S-FI junction in the P configuration. The parameters used for the calculation are tunnel resistance $R$=0.5 $\Omega$, temperature T=50 mK, $\Gamma$=0.05 $\Delta$, exchange energy $h_L$=87 $\mu$eV and $h_R$= 145 $\mu$eV for the left and right electrodes, respectively.
Blue and orange dashed lines highlight $V_1^P$ and $V_2^P$ corresponding to the two voltage threshold ($e V_{1(2)}^P = 2\Delta \mp |h_L-h_R|$). Below $V_1^P$ no current can flow through the junction. The gray-shaded area highlights the biasing range for which single-spin injection through the junction is achieved. Above $V_2^P$ the current is not spin-polarized. \textbf{c,} Tunneling current calculated in the AP configuration for the same parameters. In this configuration the energy splitting between the two spin species is increased with $e V_{1(2)}^{AP} = 2\Delta \mp |h_L+h_R|$.
By operating the valve at a fixed voltage bias $V$ within the range $V_1^{AP}\leq V \leq V_1^P$ (highlighted by the blue rectangle) a tunneling magneto resistance device with almost-perfect contrast can be implemented driving the FIs from P (\textbf{b}) to AP (\textbf{c}) configuration.}
\end{figure}
In this ideal case the device is called an {\it absolute spin valve} (ASV).   

The spectrum of a conventional  superconductors consists of a gap in the density of states (DOS) around the Fermi energy  with a large quasiparticle spectral peaks at the gap edges. By applying an external magnetic field, or by the proximity of an adjacent  magnetic insulator, the DOS of a superconducting film  shows a spin-splitting (see Fig. \ref{fig1}a). 
In the latter case the  spin-splitting is caused by an effective  exchange  field induced in the superconductor. If this field is large enough superconductivity is suppressed and the film transits to the normal state.  A switching between  the superconducting and normal state has been demonstrated recently in a device consisting of a single superconducting layer between two ferromagnetic insulators \cite{Li2013}. 

Here we report  a novel  type  of spin-valve, which shows high potential applicability as a logical switching element in a low-dissipative superconducting non-volatile memory cell.  The spin-valve is based on a tunnel junction comprising of two spin-split superconductors, as shown in Fig. \ref{fig1}b. By tuning the voltage across the junction one can obtain  very high TMR ratio  arising from the half-metallic character of the spin-split DOS.  Moreover,  at certain voltages the junction can be used as an  ideal spin source and/or spin detector  witch promises promises applicability in low-energy spintronics\cite{Giazotto2015}. 

Specifically, the valve consists of two  Al/EuS bilayers coupled through an aluminum oxide tunnel barrier (I) (see Fig. \ref{fig1}b). 
EuS is a ferromagnetic insulator (FI)  that provides a sharp spin-splitting ($\pm h$) in the DOS of the adjacent superconductor (S)  via the magnetic proximity effect \cite{Tedrow1986,Moodera1988,Miao2015},   as demonstrated in several experiments\cite{Tedrow1986,Mauger1986,Hao1990,Moodera1988,Strambini2017b}. 
This Zeeman splitting can be hundreds of  $\mu $eV to several meV, which is equivalent to that due of an external magnetic field of few to hundreds of Tesla \cite{Hao1990,Hao1991,Li2013,Tokuyasu1988} and depends on the thickness of the S layers and the quality of the FI/S interfaces.
The sign of this energy shift is opposite for each spin species, and it can be inverted by magnetizing the FI layer in the opposite direction \cite{Miao2015} (see Fig. \ref{fig1}a).

The spin valve under consideration in the present work, an FI-S-I-S-FI tunnel junction, operates between the parallel (P) and anti-parallel (AP) configuration of the two FI layers controllable by a small external magnetic field $H$, whose intensity depends on FI layer thickness and  growth conditions.
In our device the needed switching fields are a few mT \cite{Strambini2017b}.  
The working principle of the ASV can be understood from the tunneling currents flowing through the junction for the up ($+$) and down ($-$) spin species and are given by the following expressions
\begin{align}
	I_{\pm}  =  \frac{1}{eR}\int_{-\infty}^\infty & N_L(E\pm h_{L}-eV) N_R(E\pm h_{R})\times \\
   & [f_L(E-eV)-f_R(E)]dE, \,\,\,\,\,\,\,\,\,\,\,\,\,\,\,\,\,\,\,\,\,\,\,\,\,\,\,\,\,\,\,\,\,\,\,\,\,  (P) \nonumber
\label{formula1}
\end{align}

\begin{align}
	I_{\pm} =  \frac{1}{eR}\int_{- \infty}^\infty & N_L(E\mp h_{L}-eV) N_R(E\pm h_{R}) \times \\
   & [f_L(E-eV)-f_R(E)]dE, \,\,\,\,\,\,\,\,\,\,\,\,\,\,\,\,\,\,\,\,\,\,\,\,\,\,\,\,\,\,\,\,\,\,\,\,\, (AP) \nonumber
\label{formula2}
\end{align}  
for the P and AP configuration of the two FI layers. Here, $e$ is the electron charge, $f(E)$ is the Fermi-Dirac distribution function, $h_L$ and  $h_R$ are the exchange field experienced by the left and right superconductors respectively, $R$ is the normal-state resistance of the junction, $V$ the voltage bias and $N_j(E)= \left |  Re [(E+i\Gamma)/\sqrt{(E+i\Gamma)^{2}-\Delta^2}] \right |  $ the quasi-particle DOS of the $j=L,R$ left and right S layers. $\Gamma$ is the Dynes parameter that accounts for the inelastic scattering that broadens the conductance, and $\Delta$ is the superconducting pairing potential.

The above four spin-dependent currents are shown in Fig. \ref{fig1}b-c.
In the P configuration the total current is zero below a threshold voltage  $eV_1^{P} = 2 \Delta - |h_{L}-h_{R}|$  due to the absence of available states in the energy spectrum of the DOS of both spin species  (see Fig. \ref{fig1}b). This voltage threshold is decreased in the AP configuration to $eV_1^{AP}=2 \Delta - |h_{L}+ h_{R}|<eV_1^{P}$ due  to the opposite energy shift of the two DOS (see Fig. \ref{fig1}c). Thus,
by operating valve with a bias voltage $V$ between those two values, $V_1^{AP}<V<V_1^{P}$, one could achieve a spin-valve switching between a finite and near zero current flow in the AP  and P  configuration, respectively. 

The efficiency of the spin valve  can be quantified by the tunneling magneto-resistance ratio, defined as the relative difference between the junction resistances in the two configurations of the valve $TMR=(R_{P}-R_{AP})/R_{AP}$. At low temperature ($k_B T\ll \Delta$) the $TMR$ of our spin valve is only limited by the sub-gap states present in the two superconductors ($TMR \lesssim (\Delta/\Gamma)^2$). This  explains the very large value of TMR, 900$ \% $, observed in our junctions.  It is worth noticing that for voltages within this operation regime, the injected current is also 100\% spin polarized making the valve an ideal spin injector for other \textit{spintronics} applications \cite{Giazotto2008}. 

\begin{figure}
\includegraphics[width= \columnwidth]{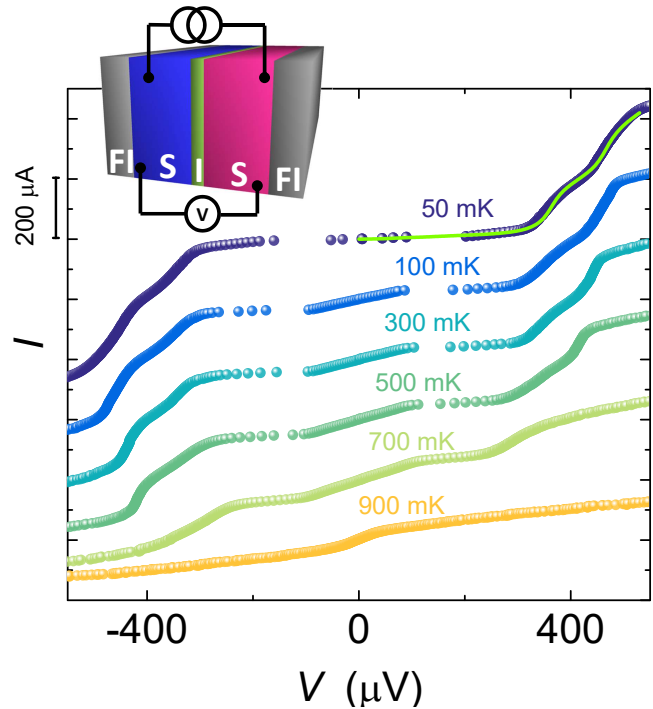}
\caption{\label{fig2} \textbf{Tunneling spectroscopy in the P configuration.} Scatter plot: $I(v)$ tunnel characteristics of an absolute spin valve at several bath temperatures after magnetic initialization of the device in the P state. Currents are plotted as a function of voltage, but data were collected setting a current bias and measuring the 4-wire voltage drop at the ends of the junction. Line plot: best-fit curve of the $I(V)$ relation (dark blue dots) at 50 mK. Data are fitted through equation 1, which yields $h_{L(R)}$  equal to 21 and 60 $\mu$eV for for the 4 and 10 nm thick EuS layers, respectively. Inset: wiring scheme of the tunneling spectroscopy setup.}
\end{figure}

The transport  properties of our devices has been explored by using a standard tunneling spectroscopy technique. The EuS/Al/Al-oxide/Al/EuS stacks are patterned into 150$\times$150 $\mu$m$^2$-wide tunnel junctions deposited by \textit{in situ} metallic shadow mask electron-beam evaporation (see Methods for fabrication details). The device tunneling spectra were obtained by measuring the current-voltage $V(I)$characteristics with four-wire contact technique using direct-current through the tunnel junctions (the biasing scheme is depicted in the inset of Fig. \ref{fig2}): the differential conductance $\frac{d I}{d V}$was evaluated via numerical differentiation of (V(I)). The spin valves were cooled in an electrical noise filtered closed-cycle He$^3$-He$^4$ dilution refrigerator equipped with a vectorial magnet which allows finely controlling the magnetization of the two FIs. All the data reported in the following are are from the same representative device. Its composition was EuS(4 nm)/Al (6.5 nm)/AlOx (tunnel barrier)/Al (6.5 nm)/EuS (10 nm).

The device was initialized by applying an in-plane 15 mT external field driving the two FI layers in the P configuration. The $V(I)$  measured at zero magnetic field after the initialization are shown in the Fig \ref{fig2} for several temperatures. As expected from Eq.~\ref{formula1}, the current exhibits a double-knee structure resulting from the mismatch between the two spin-split density of states. Although smeared, such feature persists up to more than half of the critical temperature of Al layers. By fitting the data with the tunneling current equation
$ I(V)=I_+(V) + I_-(V)$, it is possible to extract all the relevant parameters of the device, and determine the effective exchange energy induced in the S layers. 
The best fit for the $I(V)$ curves at 50 mK (green line in Fig. \ref{fig2}a) gives a $h_{L}\simeq 21 \mu$eV and $h_{R}\simeq 60 \mu$eV for the 4 nm and 10 nm thick EuS layers respectively (detailed information on the fitting procedure is described in the Methods section). These values are consistent with previously reported exchange energies in similar bilayers \cite{Strambini2017b}.

\begin{figure}
\includegraphics[width=0.9 \columnwidth]{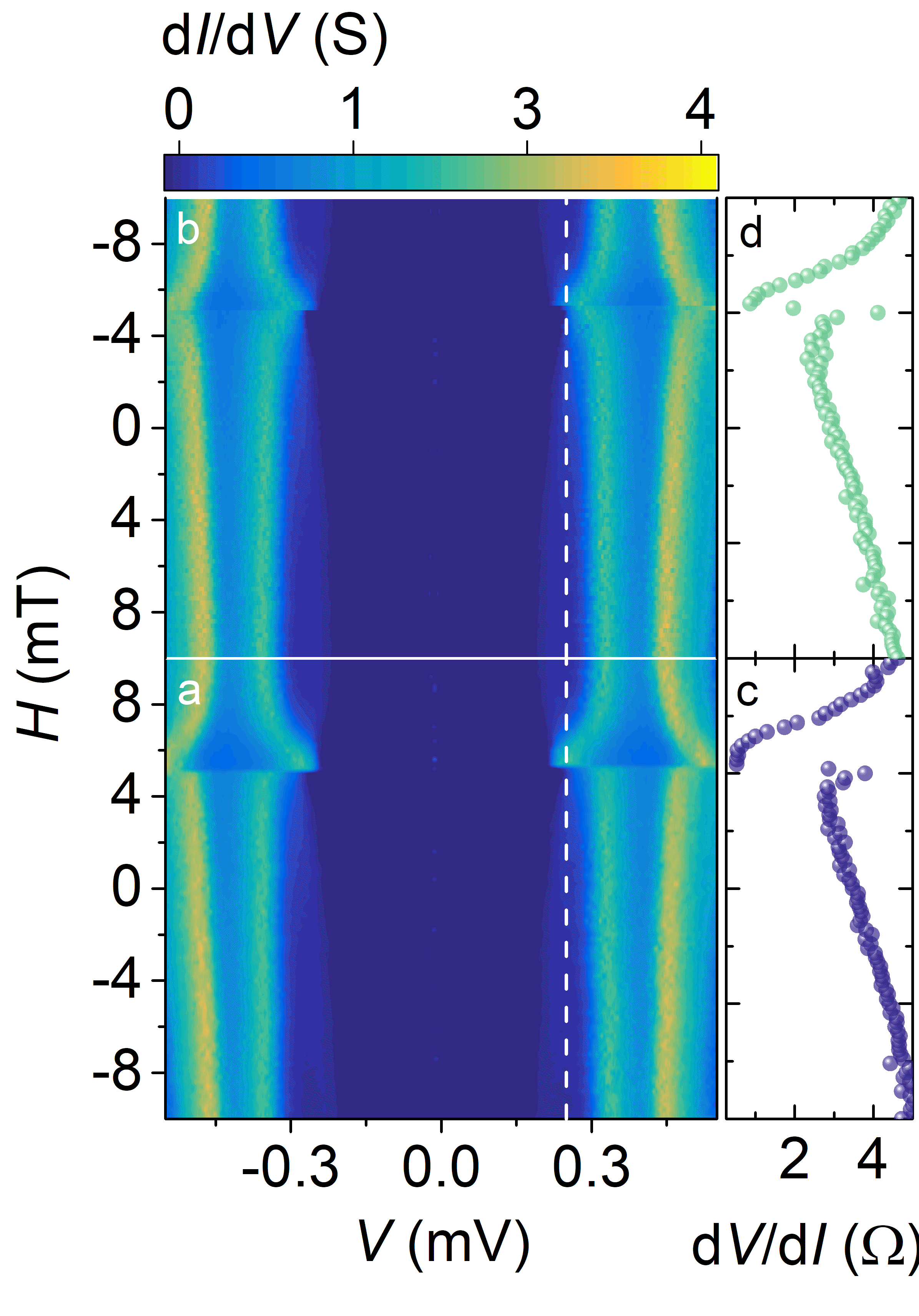}
\caption{\label{fig3} \textbf{Magnetic field evolution of the ASV.} a) and b) Color plot of the tunneling differential conductance of an ASV device at 50 mK as function of the voltage drop $V$ and of the external magnetic field $H$. Data were recorded with the same biasing scheme depicted in the inset of Fig. \ref{fig2}a. To avoid device overheating the $V(I)$ were always recorded with $I$ being swept from 0 to positive values and from 0 to negative values. $H$ was swept forward from -10 mT to 10 mT (a) and back to -10 mT (b). c) and d) Tunneling differential resistance $V = 248 \mu $V (\textit{i. e.}, along the dashed white line in panels a and b of this figure) as function of the magnetic field as $H$ was swept forward (c) and backward (d).}
\end{figure}

Importantly, the charge current flowing in the junction can  be controlled in the junction with an external in-plane magnetic field $H$ that switches the  the polarizations of the two FI layers.
Figure \ref{fig3}a-b show in color-plot the differential conductance of the ASV as a function of the voltage $V$ across the junction and $H$. Data were recorded at 50 mK by measuring the $V(I)$ tunnel characteristic of the device for each value of $H$ that was swept from -10 mT to 10 mT (Fig. \ref{fig3}a) and back to -10 mT (Fig. \ref{fig3}b). The differential conductance trace allow to distinguish three transport regimes: $1$) the \textit{suppression} regime, $2$) the \textit{spin-polarized} regime , $3$) \textit{the unpolarized} regime. 
Regime $1$ is realized in the voltage range below the voltage corresponding to the first conductance peak located at $V_{1}^P \simeq 350 \mu$ V.
Between the two conductance peaks appearing at $V_1^{P} $ and $V_2^{P} \simeq 450 \mu V$, the spin injection is maximized (regime 2) leading to a $\sim$100\% spin-polarized transport through the junction. 
The third regime occurs at higher voltages when both spin-species conduction channel are open and no preferred spin states are injected through the junction. 

To extract the different magnetic configurations, it is convenient to follow the evolution of the spin-resolved peaks at $V_1$ and $V_2$. The latter values, as discussed above, are  determined by the magnetic configuration  of the two FIs.  
Let us focus for example on Fig. \ref{fig3}a: at $H<-8$mT the system is in the P configuration. By sweeping $H$ toward positive values $V_1$ and $V_2$ remain almost constant until $H$ approaches the coercive field of the thinner FI layer, $H_4 \approx 5$mT. At that value, the magnetic polarization of this layer is inverted and a steep jump in the position of the peaks is observed. The junction then remains in the AP configuration until the coercive field $H_{10}$ of the thicker EuS layer is reached, returning the junction to the $P$ state. This second transition is smoother compared to the first one at $H_4$, as can be noted by the evolution of the split peaks between 5 and 8 mT. This is consistent with the smoother transition of the stronger 10-nm-thick FI layer~\cite{Miao2009}.

The backward sweep of $H$ (see Fig.~\ref{fig3}b) shows a similar evolution with the two coercive fields in the opposite direction confirming the ferromagnetic origin of the observed effect. 
From these data it is possible to extract the magnetoresistance of the junction at constant bias, as shown in Fig.~\ref{fig3}c and \ref{fig3}d for the forward and backward traces, respectively. 
This allows to quantify the $TMR$ of the junction, and determine the optimal operating voltage ($V_{max}$) for the ASV at which the $TMR$ is maximized.
At 50 mK, $TMR$ reaches its highest value $TMR_{max}\simeq 850\% $for $V=V_{MAX}\sim$248$\mu$V. In Fig. \ref{fig3}a-b the dashed white line shows a cut at this voltage value. This line intercepts the $V_1$ conductance peak  only in the AP configuration, while in the P state it lies within the Al sub-gap range which corresponds to a  low-conductance regime. 
It is interesting to note that this behavior is opposite to conventional magneto-resistive devices~\cite{Chappert2007}, where the the AP configuration corresponds to the  low-conductance state. 
\begin{figure*}
\includegraphics[width=0.95 \textwidth]{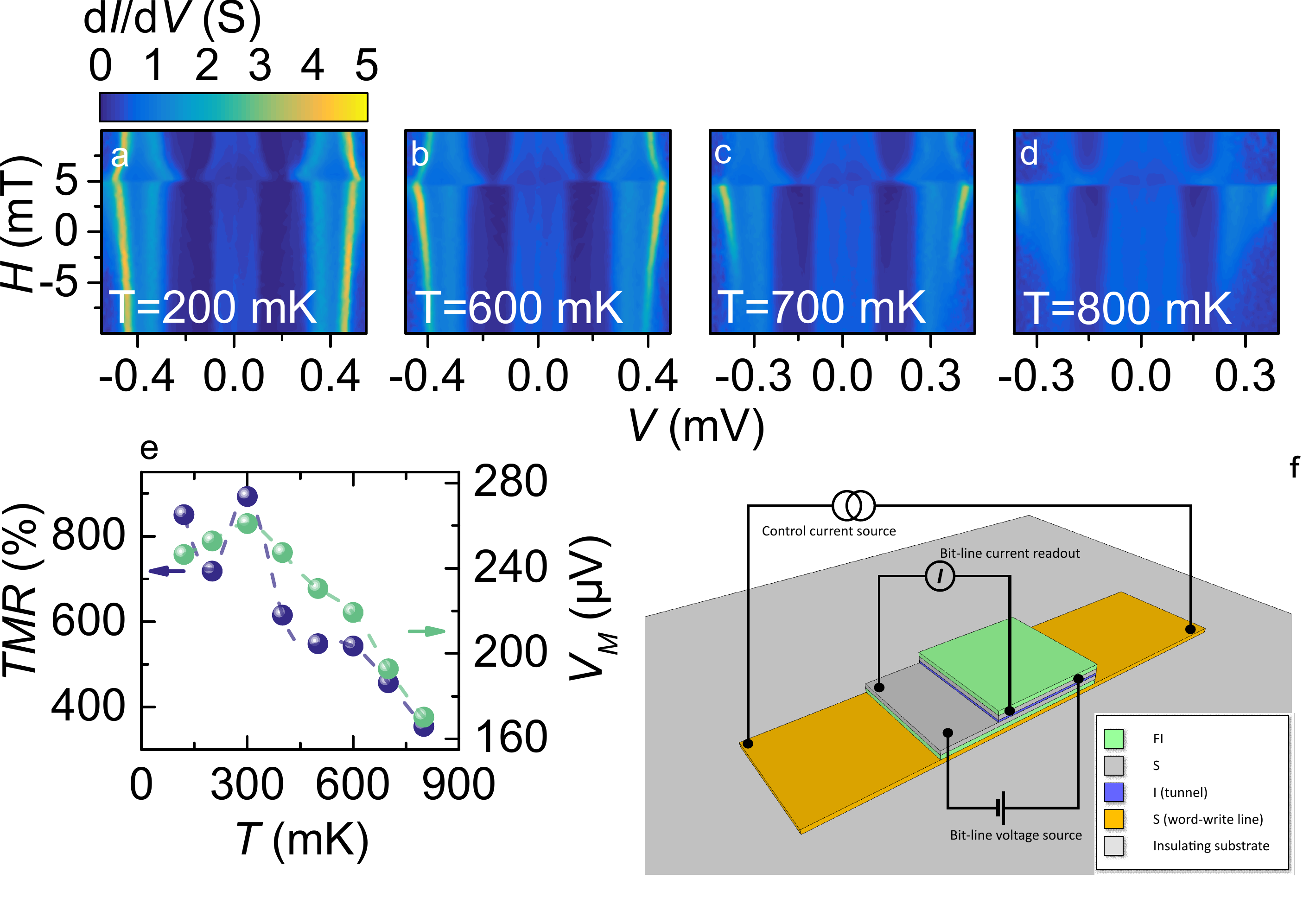}
\caption{\label{fig4} \textbf{Temperature evolution of the ASV and scheme for a memory cell.} a,b,c,d) Tunneling differential conductance of the ASV at several bath temperatures. The biasing scheme was the same shown in Fig. \ref{fig2}. e) Maximum $TMR(T)$ (left vertical scale, violet dots) as a function of temperature, which was determined by selecting for each $T$ the voltage drop $V_M$ maximizing $TMR(V)$. $V_{M}(T)$ is also plotted (right vertical scale, green dots). f) Scheme of a possible implementation of cryogenic memory cell based on the ASV.}
\end{figure*}

The temperature dependence of the $TMR$, probed by acquiring the $V(I,H)$ map at several bath temperatures $T$ is reported in Fig.~\ref{fig4}.  
This characterization demonstrate the robustness of the spin-valve effect showing a large $TMR$ up to $\sim$900 mK, \textit{i.e.} more than 75\% of the critical temperature of the Al layers, confirming the superconducting pairing potential as the main energy scale for the spin valve effect to occur in our systems.
Notably, the  maximum value of $TMR$  is not monotonic as a function of temperature, reaching its highest value of $\sim$900\% at $T$=350 mK (see Fig. \ref{fig4}e). We believe that this large $TMR$ value ranks our ASV already on par with state-of-the-art conventional spintronic devices\cite{ikeda2008tunnel,lee2007effect}. Yet, its performance can be further improved by increasing the tunnel resistance of the junction in order to reduce Joule heating and the net number of dopant impurities providing spurious leakage channels.
This could be achieved both by increasing the quality of the AlOx barrier and by reducing the spin valve lateral size. Magnetic disorder may also be detrimental in reaching a high $TMR$ ratio. Such disorder is unavoidable in polycrystalline FI layers and could be minimized by epitaxial growth of monodomain EuS layers, sharper interfaces and smaller junctions. 

The dramatic tunnel conductance difference achieved in our spin valve structure between the P and AP configuration suggests the possibility to use it as a switching element of a cryogenic memory cell. The  0- and 1-logic states can be encoded in the P and AP configurations. As shown in the magnetoresistance plots of Fig.~\ref{fig3}c and d, the readout of such a \textit{bit-line} can be  performed, \textit{e. g.}, at constant voltage-bias $V=V_1$ by measuring the tunnel current $I(V_1)$ which is expected to be zero in the P state and non-zero otherwise. The superconducting bit can be written in response to a magnetic field generated by a control current $I_{c}$ flowing in a properly designed superconducting \textit{word-write line} that controls the magnetization of the FI with the lower coercive field $H_c$ (see Fig. \ref{fig4}f).
A lower boundary of $I_{c}$ is roughly given by the relation $I_c\gtrsim H_c d/\mu_0$, where $d$ is the effective distance between the \textit{word-write} line and the FI layer, and $\mu_0$ the vacuum magnetic permeability. 
A prototype memory cell based on our ASV can have $d \simeq 4 $~nm ( \textit{i.e.}, the thickness of the first FI layer) and will need a field of $\sim$~10 mT to control the magnetization of the FI layers. This value suggests that the \textit{word-write} line has to support an $I_c \gtrsim$500$\mu$A, a current that can be injected without dissipation if the write line is made superconducting. In principle, one or both the S layers of the ASV can act as word-write line as long as the chosen superconducting material is able to support the needed current.

In summary, we have realized and demonstrated the first superconducting absolute spin-valve based on a hybrid ferromagnetic insulator/superconducting tunnel structure. The FI layers led to spin-splitting of the density of states of the adjacent superconductors through proximity induced magnetic exchange coupling. The spin valve properties of our devices were investigated by tunneling spectroscopy revealing excellent performance in terms of tunneling magneto-resistance ratio.
Our results show the ASV as a promising leading-edge building block to implement low-dissipation cryogenic nonvolatile memories\cite{Kulic2000,Baek2014,Holmes2013,Gingrich2016}. Further it has the possibility to implement a highly-polarized spin-current source for \textit{spintronic} applications\cite{Giazotto2008,Linder2015}.

\begin{acknowledgments}
We acknowledge Vitaly Golovach and Jason Robinson for fruitful discussions.  Partial financial support from the European Union's Seventh Framework Programme (FP7/2007-2013)/ERC Grant 615187-COMANCHE and FET SUPERTED are acknowledged. The work of G.D.S. is funded by Tuscany Region under the FAR-FAS 2014 project SCIADRO. JSM research at MIT is supported by NSF grant DMR 1700137 and ONR grant N00014-16-1- 2657.  The work of F.S.B  was supported by the Spanish Ministerio de Economía, Industria y Competitividad (MINEICO) under Projects No. FIS2014-55987-P and FIS2017-82804-P. 
\end{acknowledgments}

\section{AUTHOR CONTRIBUTIONS}
E.S., F.G, J.S.M. and F.S.B proposed and designed the experiment. G.D.S.  wrote the paper performed the measurements and analyzed the experimental data with input from F.G. J.S.M. fabricated the samples. All authors discussed the results and their implications equally at all stages.


\section*{Methods}

\subsection{Sample fabrication and measurement}

Samples consist of cross bars fabricated by in situ metallic shadow mask electron-beam evaporation, defining a tunnel junction. The investigated prototype device is based on the following material stack: EuS(4)/Al(6.5)/Al$_2$O$_3$/Al(6.5)/EuS(10) (thickness in nanometers, listed in the order in which they were deposited). The junctions were fabricated in a vacuum chamber with a base pressure $2\times 10^{-8}$ Torr. To facilitate the growth of smooth films, a thin Al$_2$O$_3$ (1 nm) seed layer was deposited onto substrate.
The junctions were capped with 4 nm of Al$_2$O$_3$ for protection. The device were wedge-bonded on standard \textit{dual-in-line} chip carriers and cooled down in He$^3$/He$^4$ closed-cycle dilution refrigerator. For all the measurements reported in this paper the magnetic field was applied in the plane of the junction oriented at $\sim$45$^\circ$ with respect to the axes defined by device cross. $TMR$ ratios of the spin valves were also probed as a function of such angle with no appreciable differences with respect to the data shown in the paper. 

\subsection{$V(I)$ fit procedure and results.}
 The $V(I)$ characteristic at 50 mK was fitted by using Eq. 1. The parameters left free to vary during the fitting procedure and the results obtained are resumed in table \ref{tab:table2}. The coefficient of determination of the fit ($R^2$) is 0.994. 
\begin{table}
	\begin{ruledtabular}
		\begin{tabular}{ccc}
			Paremeter & Fit Value & Description\\
			\hline
			$R$ & 0.5313 $\Omega$ & Tunnel resistance\\
			$T$ & 0.2276 K & Electronic temperature\\
			$\gamma$ & 0.0496 & Dynes parameter\\
			$H_L$ & 0.356 T & Exchange fields $h_{L}/\mu_B$ \\
			$H_{R}$ & 1.044 T & and Exchange field $h_{R}/\mu_B$\\
		\end{tabular}
	\end{ruledtabular}
	\caption{\label{tab:table2} Result of the fit procedure on the $V(I)$ characteristic of the ASV at 50 mK.}
\end{table}
It is worth to note that the extracted electronic temperature is substantially higher than the cryostat bath temperature which could be ascribed to quasi-particle overheating due to the weak electron-phonon thermal coupling existing at ultra-low temperatures.

\bibliographystyle{apsrev4-1}
\bibliography{references}

\end{document}